# Control of Valley Degeneracy in MoS$_2$ by Layer Thickness and Electric Field and Its Effect on Thermoelectric Properties

Jisook Hong[1], Changhoon Lee[1], Jin-Seong Park[2] and Ji Hoon Shim[1,3,*]

[1] Department of Chemistry, Pohang University of Science and Technology, Pohang 790-784, Korea

[2] Division of Materials Science and Engineering, Hanyang University, Seoul 133-719, Korea

[3] Department of Physics and Division of Advanced Nuclear Engineering, Pohang University of Science and Technology, Pohang 790-784, Korea

## ABSTRACT

We have investigated the valley degeneracy of MoS$_2$ multilayers and its effect on thermoelectric properties. By modulating the layer thickness and external electric field, the hole valleys at Γ and K points in the highest energy valence band and the electron valleys at K and Σ$_{min}$ points in the lowest energy conduction band are shifted differently. The hole valley degeneracy is observed in MoS$_2$ monolayer, while that of electron valley is in MoS$_2$ bilayer and monolayer under the external electric field. By tuning the valley degeneracy, the Seebeck coefficient and electrical conductivity can be separately controlled, and the maximum power factor can be obtained in n-type (p-type) MoS$_2$ monolayer with (without) the external electric field. We suggest that the transition metal dichalcogenides are good example to investigate the role of valley degeneracy in the thermoelectric and optical properties with the control of interlayer interaction and external electric field strength.

I. INTRODUCTION

During the last half century, layered transition-metal dichalcogenides (TMDCs) have attracted much attention because of their characteristic quasi two-dimensional structure and anomalous physical properties such as superconductivity[1–3] and charge density wave.[4–6] Recently, as the exfoliation technique makes it possible to obtain the monolayers of two-dimensional materials such as graphene[7], the layered TMDCs have been in the spotlight again. Especially, 2H-$MoS_2$ and its family such as $MoSe_2$, $WS_2$ and $WSe_2$ have been studied a lot because their indirect band gap in bulk changes into direct band gap in their monolayer limit.[8–12] Because of its large direct band gap and two dimensionality, TMDCs monolayer has been applied to nano-electronics[13–15] and nano-optoelectronics.[16–18]

TMDCs also have been studied as candidates for thermoelectric materials. It is known that semiconducting TMDCs have high Seebeck coefficient ($S$) (700-900 µV/K)[19,20] because of their characteristic quasi two-dimensional crystal structures. Therefore, TMDCs have a possibility of high thermoelectric efficiency which is described by the dimensionless figure of merit $ZT=S^2\sigma T/\kappa$, where $\sigma$, $T$ and $\kappa$ represent the electrical conductivity, the operating temperature and thermal conductivity. However, the electrical conductivity of $MoS_2$ and its family is usually very low[19,20] and therefore $MoS_2$ has been reported to have low thermoelectric efficiency $ZT$ below 0.1.[21] In order to improve their thermoelectric performance, it is essential to enhance their electrical conductivity while keeping the high Seebeck coefficient. There have been a few theoretical studies to enhance the thermoelectric efficiency of TMDCs by increasing their electrical conductivity. Guo *et al.* investigated the pressure effect on the electronic structure and transport properties of $MoS_2$. They predicted that $ZT$ of $MoS_2$ can be enhanced up to 0.65 under high pressure.[22] It was also suggested that the electrical conductivity and the power factor (PF) of TMDCs can be increased by making hypothetic mixed layer system such as $MoS_2/MoTe_2$ or $WS_2/WTe_2$.[23,24] In these studies, they mainly focused on the increase of the electrical conductivity by modifying the band gap using the interlayer interactions. Also, Wickramaratne *et al*. and Huang *et al*. investigated the thermoelectric properties of multilayer TMDCs depending on their thickness by using density functional theory (DFT) calculations[25,26].

In this work, we have focused on the control of the valley degeneracy in $MoS_2$ to improve thermoelectric properties. In two bands system, total electrical conductivity and Seebeck coefficient can be described by

$$\sigma_{total} = \sigma_1 + \sigma_2 \tag{1}$$

$$S_{total} = \frac{\sigma_1 S_1 + \sigma_2 S_2}{\sigma_{total}}, \tag{2}$$

where the subscript means the band index. If we assume that two bands are degenerate and their contribution to the electrical conductivity is same, the total electrical conductivity is enhanced by twice while the total Seebeck coefficient remains the same. This band convergence approach has been applied to various thermoelectric materials to enhance the PF.[27,28] In the band structures of $MoS_2$, we identified that there are several hole/electron valleys near the valence/conduction band edge and their degeneracy can be severely affected by its thickness or the external electric field. We also showed that the converged valleys enhanced the electrical conductivity and the PF of $MoS_2$ system. We suggest that the thermoelectric properties of $MoS_2$ should be optimized by controlling valleys near the band gap.

## II. CALCULATION DETAILS

In the calculation of the electronic structures, we used the experimental crystal structure of 2H-MoS$_2$ which has the space group *P6$_3$/mmc* with the cell parameters of *a*=3.169 Å and *c*=12.324 Å. In the unit cell, Mo is at (1/3, 2/3, 1/4) and S is at (1/3, 2/3, *u*) where *u* is 0.623.[29] For the multilayer structures, we considered bulk, 4-layers, 2-layers (bilayer) and 1-layer (monolayer) MoS$_2$ structures with enough vacuum between separated multilayers.

In the first-principles calculation, we employed the full potential linearized augmented plane-wave (LAPW) method encoded in WIEN2k code[30] under the generalized gradient approximation (GGA) of Perdew-Burke-Ernzerhof (PBE)[31] using 300 k-points in the full Brillouin zone. We did not consider the spin-orbit coupling because the spin-orbit splitting does not give significant effect on the valley degeneracy or transport properties. The effect of perpendicular external electric field has been considered in 1-layer (monolayer) MoS$_2$ with the size of 0.11, 0.22, 0.33, 0.44 and 0.55 eV/Å. We also compared the electronic structures of 1-layer MoS$_2$ using frozen-core projector augmented-wave (PAW) method as implemented in the Vienna *ab initio* simulation package (VASP).[32,33] Consistent results were obtained for two sets of calculations in the band structures.

Transport properties were calculated using BoltzTraP code[34] which is based on Boltzmann transport theory under the rigid band approximation and constant relaxation time approximation. Guo *et al*. demonstrated that this approach described well the experimental

Seebeck coefficient and electrical conductivity of bulk $MoS_2$.[21] To make sure the convergence of the calculation, we used dense 18000 k-points in the full Brillouin zone.

### III. RESULTS AND DISCUSSION

Accurate description of the band gap is usually crucial for the transport properties of semiconductor because thermally excited charge carriers mainly contribute to the electrical conductivity. We calculated the band structures of $MoS_2$ using PBE-GGA and EV-GGA which estimate the band gaps of 0.883 and 1.039 eV, respectively. Although the band gap from EV-GGA is much closer to the experimental one (1.23 eV[35]), there is no significant difference in the Seebeck coefficient and electrical conductivity at the carrier concentration above ~ $10^{17}$ /cm$^3$. (See Fig. S1) At high carrier concentration, the transport properties are mainly contributed by doped charge carriers, which are dominated by the band dispersion and scattering rate, instead of the excited carriers. Because we are interested in high carrier concentration region for the optimal thermoelectric properties, we make a discussion with the results from PBE-GGA method.

#### A. Control of valley degeneracy by thickness or external electric field

Figure 1 (a) shows the calculated band structures of $MoS_2$ as varying the layer thickness from bulk (infinite thickness) to 1-layer. Our results are in good agreement with the reported electronic structures of bulk and multilayers of $MoS_2$.[9–12] The highest valence band has hole valleys at Γ and K points and the lowest conduction band also has electron valleys at $\Sigma_{min}$ and K points, where $\Sigma_{min}$ locates between Γ and K points. In bulk $MoS_2$, CBM and VBM locate at Γ and $\Sigma_{min}$ points, respectively. So, it shows an indirect band gap of 0.883 eV. As reducing the layer thickness, the conduction band at $\Sigma_{min}$ shifts upward, while the valence band at Γ shifts downward, continuously. Because the bands at K point are rather robust to the change of layer thickness, 1-layer $MoS_2$ shows the direct band gap feature at K point with the band gap size of 1.757 eV. Interestingly, the convergence of valley is shown during the evolution of layer thickness. The constant energy surfaces near the band gap edges indicate clearly the valley degeneracy as shown in Figs. 1 (b) and (c). In the valence band, bulk $MoS_2$ does not have valley degeneracy, while 1-layer $MoS_2$ has nearly degenerate hole valleys at Γ and K points and there are three cylindrical constant energy surfaces in the Brillouin zone. In the conduction band, bulk $MoS_2$ has 6-fold valley degeneracy at $\Sigma_{min}$ points, and 1-layer $MoS_2$ has 2-fold valley degeneracy at K points. Note that 2-layers $MoS_2$ shows the convergence of electron valleys at K and $\Sigma_{min}$ points, thus there is ten-fold valley degeneracy. At the convergence of the valleys, the density of states is much enhanced at the band edge and so the charge carriers are expected to be increased.

The different behavior of the valleys with the layer thickness can be understood from the different hybridization strength between Mo 4$d$-orbital and S 3$p$-orbital in each valleys. The hole/electron valleys at K point are mainly contributed from Mo 4$d$-orbital which is localized between neighboring S layers. (See Fig. S2) Thus the energy levels of the hole/electron valleys at K point are not sensitive to the layer thickness or the number of neighboring layers. However, the hole/electron valleys at Γ and $\Sigma_{min}$ points are contributed by the strongly mixed orbital of both S 3$p$-orbital and Mo 4$d$-orbital. Therefore, the energy levels of hole/electron valleys at Γ/$\Sigma_{min}$ points are sensitively shifted with the change of interlayer interaction.

The energy levels of valleys at Γ/$\Sigma_{min}$ points are also adjustable by applying external electric field as done by the control of the layer thickness. Figure 2 shows the band structure of 1-layer MoS$_2$ depending on the electric field strength. The hole valley at Γ point and electron valleys at Σ$_{min}$ point are shifted downward as increasing the electric field whereas the hole and electron valleys at K point remain robust against the external field. So, the valley degeneracy of valence band at K and Γ points in 1-layer MoS$_2$ is lifted and the VBM is shown only at K point under the electric field. At the conduction band, the valley at Σ$_{min}$ point shift downward and the convergence of valleys at K and Σ$_{min}$ points are observed around 0.55 V/Å. With further increased electric field, the Σ$_{min}$ point is expected to be the CBM. Note that the band gap properties are also changed with the modulation of valley degeneracy. The indirect band gap from Γ to Σ$_{min}$ points in bulk MoS$_2$ becomes the direct band gap at K point in 1-layer MoS$_2$. With the electric field above 0.55 V/Å on 1-layer MoS$_2$, the indirect band gap feature reappears at different momentum from K to Σ$_{min}$ points. It will be interesting to investigate the evolution of optical properties under the control of the layer thickness and electric field.

The change of the band structures in MoS$_2$ under electric field can be understood from the change of hybridization strength with the asymmetric charge distribution. In Fig. 3, we plot the absolute value of wave function $|\psi(\mathbf{r})|$ of 1-layer MoS$_2$ without and with the external electric field. All the bands near the band gap are mainly contributed from Mo 4$d$-orbital and it agrees well with the high charge density around Mo atoms. The VBM at Γ point and CBM at Σ$_{min}$ point, however, have relatively large contribution of S 3$p$-orbital compared with the VBM and CBM at K point (Fig. S2). The charge density of hole valley at Γ point and electron valley at Σ$_{min}$ point are affected much by the external electric field as shown in Fig. 3. When there is applied electric field along the out of plane direction, the energy levels of upper S layer and lower S layer are split and therefore the energy level of anti-bonding states are lowered. The bands with the localized Mo 4*d*-orbital are not affected by the electric field because they are nearly non-bonding states. As a result, only the hole valley at Γ point and electron valley at $\Sigma_{min}$ point are shifted downward under the electric field.

**B. Thermoelectric properties depending on valley degeneracy**

The thermoelectric properties of $MoS_2$ multilayers are calculated along the in-plane direction. Figure 4 shows the Seebeck coefficient, electrical conductivity, and PF of $MoS_2$ multilayers at 300 K. The chemical potential is represented with respect to the VBM and CBM in p-type and n-type cases, respectively. The Seebeck coefficients are not affected so much by the change of layer thickness. Small change in Seebeck coefficient can be understood from the averaged contribution from different valleys as indicated by Eq. (2). Dimensional dependency is not significant near the band edge because bulk $MoS_2$ already has 2-dimensional electronic structures with weak interlayer interaction. However, the electrical conductivity shows clear changes depending on the layer thickness. Among p-type $MoS_2$ multilayers, 1-layer $MoS_2$ shows significantly increased electrical conductivity as well as the PF. This result is mainly caused by the valley degeneracy at the valence band edge. Similarly, among n-type $MoS_2$ multilayers, 2-layers $MoS_2$ with the converged electron valleys has the highest electrical conductivity while maintaining high Seebeck coefficients. Therefore, 1-layer $MoS_2$ and 2-layers $MoS_2$ have the highest power factors when the corresponding carrier densities are $6.38 \times 10^{20}$ cm$^{-3}$ for p-type and $5.44 \times 10^{20}$ cm$^{-3}$ for n-type, respectively.

The maximum peak values of the PF near the band edge as varying the temperature are shown in the inset of Figs. 4 (e) and (f). In the whole temperature range from 300 K to 700 K, 1-layer $MoS_2$ and 2-layers $MoS_2$ have the best power factors among p-type and n-type

MoS$_2$ multiplayers, respectively. In previous theoretical studies, it was reported that the maximum *ZT* of p-type bulk MoS$_2$ is predicted to be 0.1 at 700 K with the carrier concentration of ~ $10^{20}$ cm$^{-3}$.[21] According to our calculation, we suggest that six times higher *ZT* can be achieved in p-type 1-layer MoS$_2$ if we assume there is no significant changes of thermal conductivity *κ*.

The thermoelectric properties of 1-layer MoS$_2$ under electric field were also explored and summarized in Fig. 5. Seebeck coefficient of 1-layer MoS$_2$ remains almost same (Figs. 5 (a) and (b)) and the electrical conductivity changes significantly (Figs. 5 (c) and (d)) with the modulation of valley degeneracy under the electric field strength. Similar to the study of the layer thickness, the electrical conductivity has the highest value when there is valley degeneracy, namely, p-type without external electric field and n-type under the field around 0.55 V/Å. The PFs of 1-layer MoS$_2$ in the absence of electric field show multiple peaks because of the contribution from K and $\Sigma_{min}$ valleys with different energy levels. As those two valleys converge, the PF peaks converge near the band edge and the maximum PFs can be obtained. (Figs. 5 (e) and (f)) In the case of hole doping (p-type), the valley degeneracy becomes broken with the electric field, so the PFs are suppressed. In real applications, the effect of substrates or molecule absorption on the surface would be similar to that of external field on the layered TMDCs. It could be interesting to investigate the role of electric field on the transport and thermoelectric properties in other two-dimensional materials.

In Fig. 6, we summarize all the discussion on the control of valley. Because of the different orbital characters, the energy level of valley is controllable by modulating multilayer thickness or external electric field. Only the valleys at Γ and $\Sigma_{min}$ points are shifted upward or downward while the valleys at K point keep the almost same energy level. So, the size of the direct band gap at K point is rather insensitive to the layer thickness and electric field, and the change of band gap is mainly induced by the hole/electron valleys at $\Gamma$ and $\Sigma_{min}$ points. Also the valley degeneracy contributes greatly on optimizing the thermoelectric performance. Our results show that the thermoelectric properties are optimized when there is valley degeneracy.

## IV. CONCLUSIONS

Using the *ab-initio* calculation, we have investigated the thermoelectric properties of $MoS_2$ depending on the valley degeneracy near the band gap. The valley convergence is modulated by changing the multilayer thickness and applying perpendicular electric field. By tuning the valley degeneracy, the electrical conductivity is enhanced while the Seebeck coefficient is not affected significantly. Also, three is a clear change of the band gap properties depending on the modulation of the valley degeneracy. Therefore we suggest that $MoS_2$ and other TMDC can be used for tuning the thermoelectric and optical properties by the valley degeneracy.

## ACKNOWLEDGMENT

The authors thank Moon Ho Cho for fruitful discussion. This research is supported by Global Frontier Program "Global Frontier Hybrid Interface Materials" (2013M3A6B1078870).

Corresponding Author

*E-mail: jhshim@postech.ac.kr

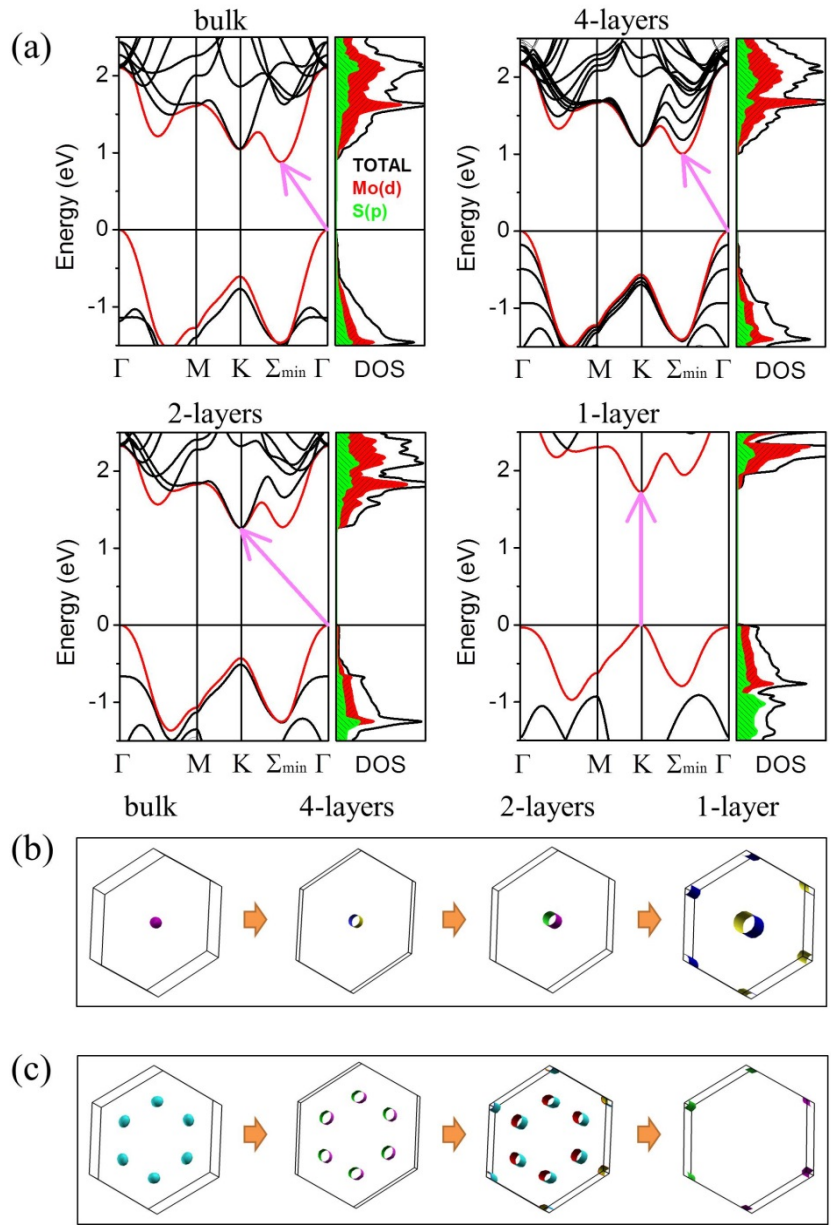

**Figure 1** (Color online) (a) Band structures and density of states of bulk, 4-layers, 2-layers and 1-layer $MoS_2$. Pink arrows represent the direct/indirect band gaps. (b), (c) The constant energy surfaces in the Brillouin zone where the energy levels are below/above the valence/conduction band edge by 0.06 eV. The number of electron/hole cylinders in the Brillouin zone changes depending on the thickness of the $MoS_2$ multilayer.

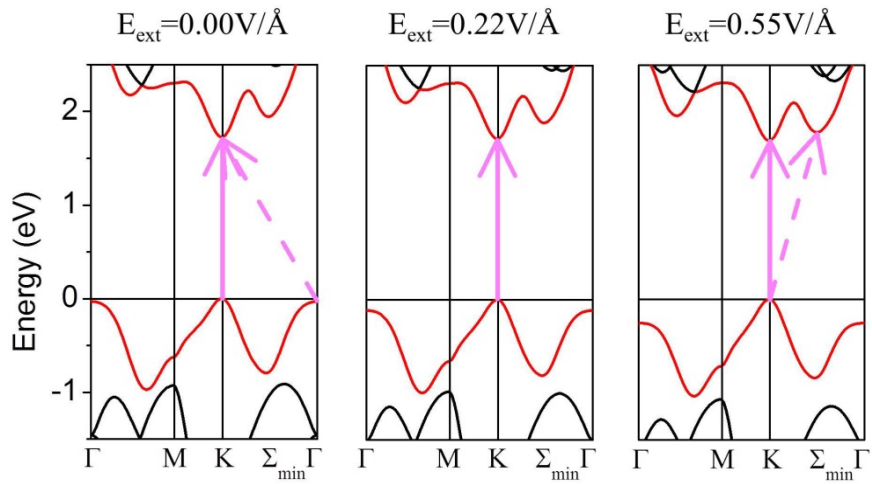

**Figure 2** (Color online) Band structures of 1-layer $MoS_2$ under the electric field of 0, 0.22 and 0.55V/Å. The hole valleys at Γ and K points nearly converge without electric field and the electron valleys at $\Sigma_{min}$ and K points do at E=0.55 V/Å. Pink arrows indicate the direct/indirect band gaps.

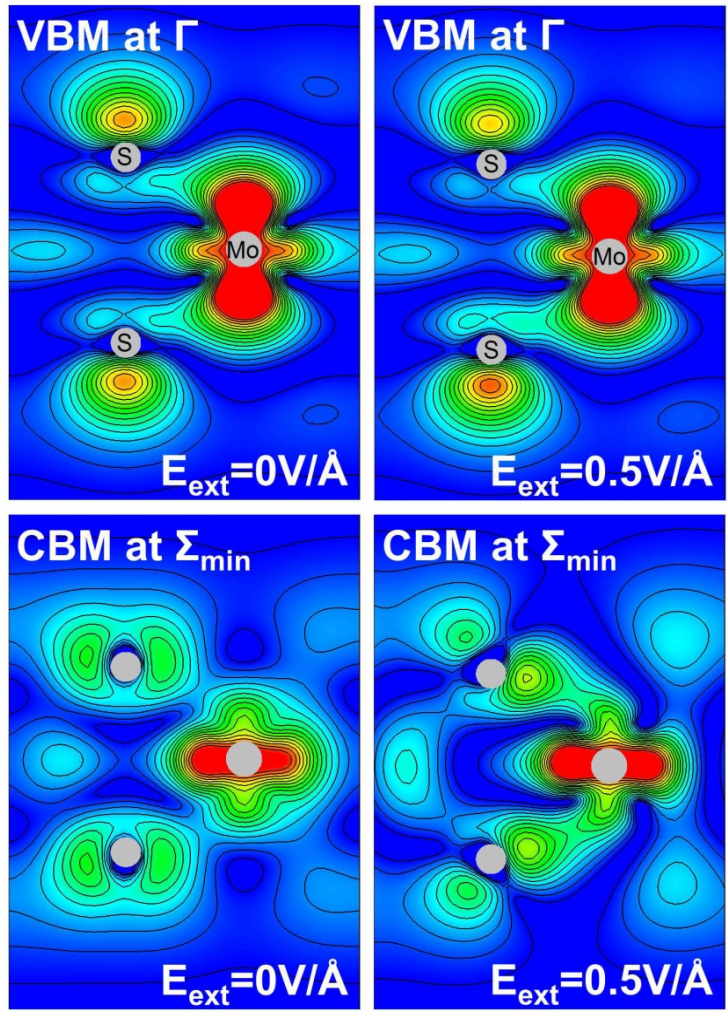

**Figure 3** (Color) The absolute value of wave function $|\psi(\mathbf{r})|$ of 1-layer MoS$_2$. Red and blue represent dense and rare charge density, respectively. Grey circles are Mo and S atoms in the unit cell.

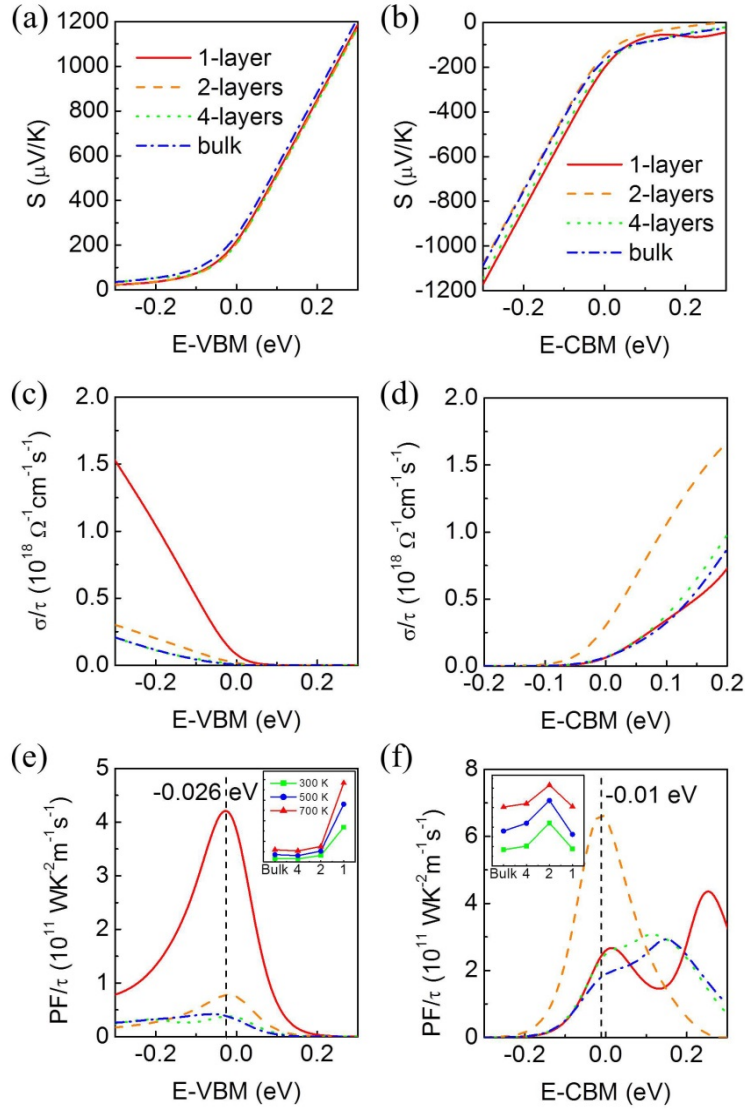

**Figure 4** (Color online) In-plane Seebeck coefficient(S), electrical conductivity($\sigma/\tau$) and power factor(PF/$\tau$) of MoS$_2$ at 300 K are plotted depending on their thickness and chemical potential. The left column ((a), (c) and (e)) is the property of p-type MoS$_2$ and the right column ((b), (d) and (f)) is that of n-type MoS$_2$. The chemical potential 0.0 refers to the valence and conduction band edges for p- and n-types, respectively. The insets of (e) and (f) show the maximum power factor depending on the temperature.

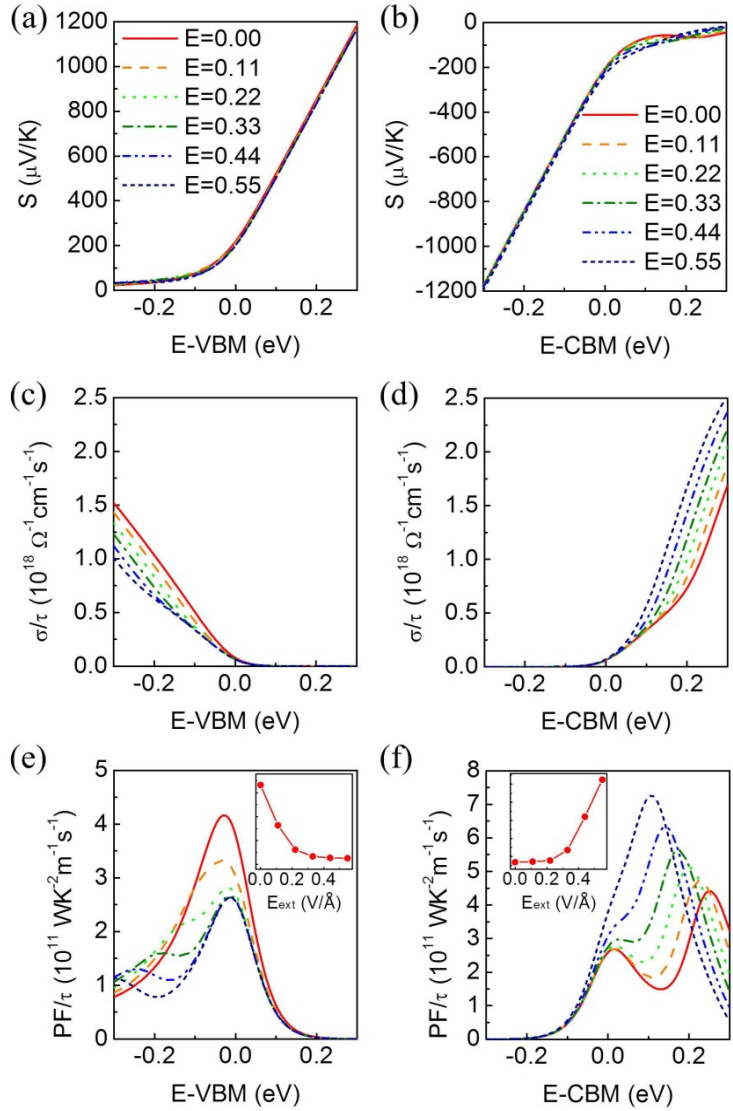

**Figure 5** (Color online) In-plane Seebeck coefficient, electrical conductivity and power factor of 1-layer $MoS_2$ at 300 K are plotted depending on the external electric field strength and chemical potential. The left column ((a), (c) and (e)) is the property of p-type 1-layer $MoS_2$ and the right column ((b), (d) and (f)) is that of n-type 1-layer $MoS_2$. Near the band edge, p-type (n-type) 1-layer $MoS_2$ has maximum power factor when the electric field E=0 V/Å (E=0.55 V/Å). Inset of (e) and (f) is the maximum power factor depending on the electric field strength. It is clearly shown that the power factors of p- and n-type have the opposite trend and p-type has the maximum under E=0 V/Å whereas n-type does under 0.55 V/Å.

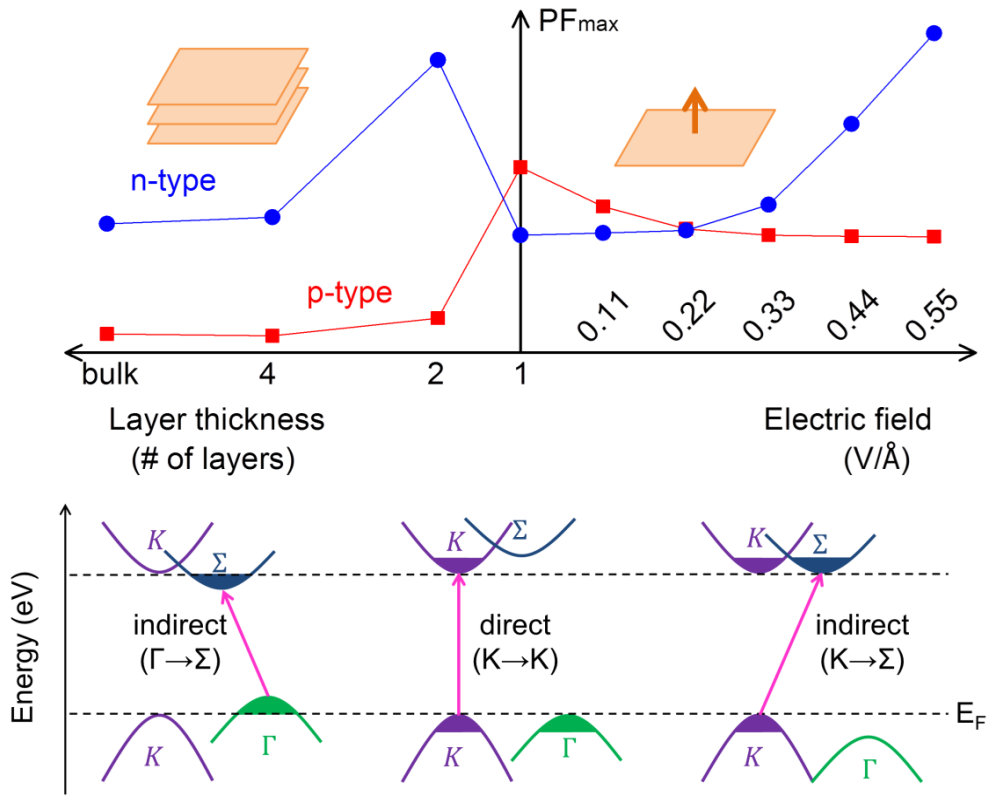

**Figure 6** (Color online) Summary of valley modulation depending on multilayer thickness and external field and its effect on thermoelectric efficiency.

# Supporting Information

**Control of Valley Degeneracy in MoS$_2$ by Layer Thickness and Electric Field and Its Effect on Thermoelectric Properties**

Jisook Hong[1], Changhoon Lee[1], Jin-Seong Park[2] and Ji Hoon Shim[1,3,*]

[1] Department of Chemistry, Pohang University of Science and Technology, Pohang 790-784, Korea

[2] Division of Materials Science and Engineering, Hanyang University, Seoul 133-719, Korea

[3] Department of Physics and Division of Advanced Nuclear Engineering, Pohang University of Science and Technology, Pohang 790-784, Korea

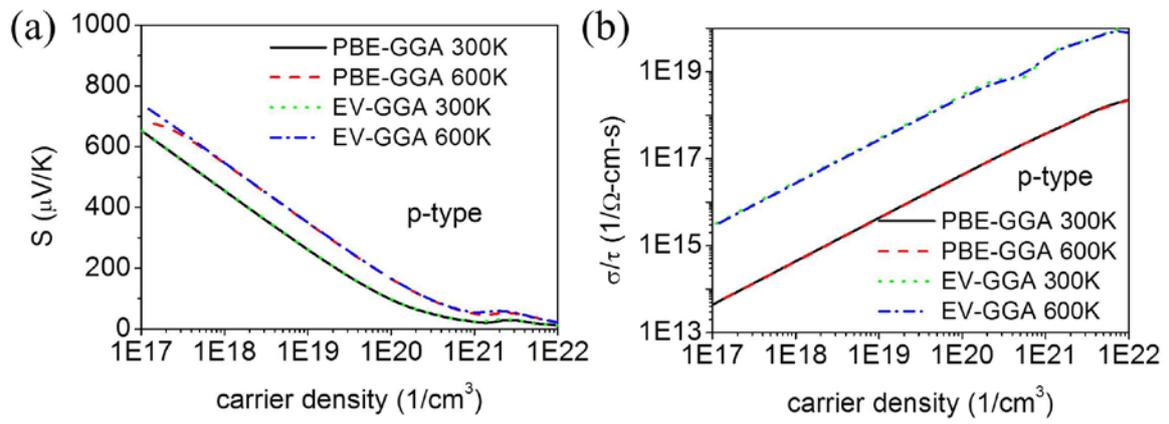

**Figure S. 1** (Color online) Comparison between calculated (a) Seebeck coefficients and (b) electrical conductivity using electronic structure from PBE-GGA and EV-GGA.

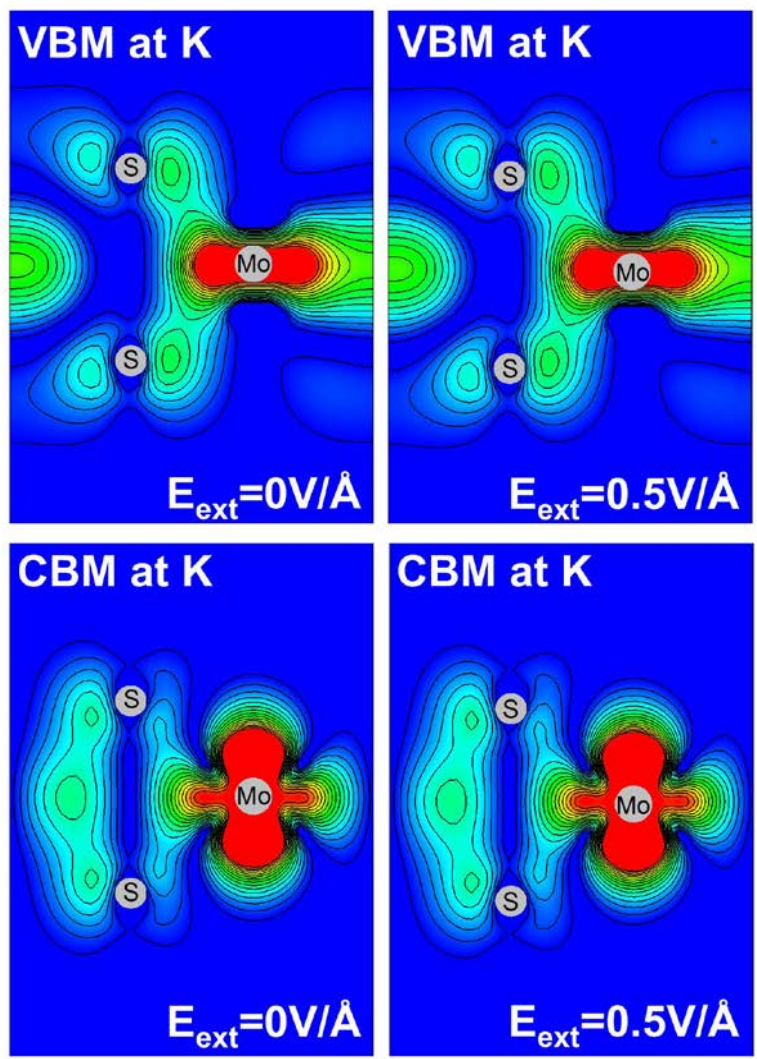

**Figure S. 2** (Color) The absolute value of wave function $|\psi(\mathbf{r})|$ of 1-layer MoS$_2$. Red and blue represent dense and rare charge density, respectively. Grey circles are Mo and S atoms in the unit cell.